\documentclass[useAMS,usenatbib]{mn2e}

\usepackage{graphicx}

\title[Optical timing of the Crab pulsar with Iqueye]{Optical phase coherent timing of the Crab nebula pulsar with Iqueye at the ESO New Technology Telescope}
\author[L. Zampieri et al.]{L. Zampieri$^{1}$\thanks{E-mail: luca.zampieri@oapd.inaf.it}, A. \v{C}ade\v{z}$^{2}$, C. Barbieri$^{3}$, G. Naletto$^{4,5}$, M. Calvani$^{1}$, M. Barbieri$^{1,3}$, \newauthor
E. Verroi$^{3,4}$, P. Zoccarato$^{6}$, T. Occhipinti$^{7}$
\\
$^{1}$INAF-Astronomical Observatory of Padova, Padova, 35122, Italy \\
$^{2}$Faculty of Mathematics and Physics, University of Ljubljana, Ljubljana, 1000, Slovenia \\
$^{3}$Department of Physics and Astronomy, University of Padova, Padova, 35131, Italy \\
$^{4}$Department of Information Engineering, University of Padova, Padova, 35131, Italy \\
$^{5}$CNR-IFN UOS Padova LUXOR, Padova, 35131, Italy \\
$^{6}$Trimble Terrasat GmbH, H\"ohenkirchen-Siegertsbrunn, D-85635 Munich, Germany \\
$^{7}$Adaptica s.r.l., Padova, 35129, Italy
}
\begin{document}

\date{Accepted ... Received ...; in original form ...}

\pagerange{\pageref{firstpage}--\pageref{lastpage}} \pubyear{2013}

\maketitle

\label{firstpage}

\begin{abstract}
The Crab nebula pulsar was observed in 2009 January and December with a novel very fast optical photon counter, Iqueye, mounted at the ESO 3.5 m New Technology Telescope. Thanks to the exquisite quality of the Iqueye data, we computed accurate phase coherent timing solutions for the two observing runs and over the entire year 2009. Our statistical uncertainty on the determination of the phase of the main pulse and the rotational period of the pulsar for short (a few days) time intervals are $\approx 1 \, \mu$s and $\sim$0.5 ps, respectively. Comparison with the Jodrell Bank radio ephemerides shows that the optical pulse leads the radio one by $\sim 240 \, \mu$s in January and $\sim 160 \, \mu$s in December, in agreement with a number of other measurements performed after 1996. A third-order polynomial fit adequately describes the spin-down for the 2009 January plus December optical observations.
The phase noise is consistent with being Gaussian distributed with a dispersion $\sigma$ of $\approx 15 \, \mu$s in most observations, in agreement with theoretical expectations for photon noise-induced phase variability.
\end{abstract}


\section{Introduction}
\label{sec1}

Only a handful of pulsars show optical emission at detectable levels and, among them, only five show pulsations in the optical band: the Crab (PSR B0531+21; PSR J0534+2200) and Vela (PSR B0833-45) pulsars, PSR B0540-69, PSR B0656+14, and Geminga (PSR B0630+17) (see e.g. \citealt{mignani11}). The first source in which such pulsations were detected is the Crab nebula pulsar \citep{cocke+69,lynds+69}. As such, the study of these pulsars, and in particular of the Crab which is the brightest among them ($V\sim 16.6$ mag), is of noticeable importance to understand the optical emission mechanism and to investigate, by comparison with the radio and other wavebands, the geometry of the emission regions and of the magnetic field.

\begin{table*}
 \centering
 \begin{minipage}{140mm}
  \caption{Log of the 2009 observations of the Crab nebula pulsar, taken with Iqueye 
  mounted at the ESO 3.5 m NTT telescope in Chile. Times refer to the barycentre
  of the Solar system (TEMPO1).}
  \label{tab1}
  \begin{tabular}{@{}llrrrrlrlr@{}}
  \hline
    &  Observation ID  &  Start time (UTC)  & Start time (MJD) & Duration (s) \\
 \hline
            1 & 20090115-030226UTC$_{-}$crab2 & January 15, 03:11:14.9 & 54846.132811 &  255.2  \\
            2 & 20090115-031933UTC$_{-}$crab3 & January 15, 03:28:07.8 & 54846.144535 &  198.2  \\
            3 & 20090115-032345UTC$_{-}$crab4 & January 15, 03:32:18.8 & 54846.147440 &  897.8  \\            
            4 & 20090116-014452UTC$_{-}$crab & January 16, 01:53:40.9 & 54847.078946 &  156.9  \\
            5 & 20090116-014802UTC$_{-}$crab & January 16, 01:56:48.7 & 54847.081119 &   98.2  \\
            6 & 20090116-015011UTC$_{-}$crab & January 16, 01:59:10.9 & 54847.082765 &  233.7  \\
            7 & 20090116-015448UTC$_{-}$crab2 & January 16, 02:03:37.6 & 54847.085852 &  1997.6 \\
            8 & 20090117-012137UTC$_{-}$crab & January 17, 01:30:18.3 & 54848.062712 &  3597.5  \\
            9 & 20090118-014557UTC$_{-}$crab & January 18, 01:55:02.2 & 54849.079887 &  727.1   \\
           10 & 20090118-020018UTC$_{-}$crab & January 18, 02:09:22.3 & 54849.089842 & 5997.2   \\
		   11 & 20090119-004736UTC$_{-}$crab & January 19, 00:56:09.6 & 54850.039000 &  3598.1	\\
           12 & 20090120-012735UTC$_{-}$crab & January 20, 01:36:16.4 & 54851.066857 &  3712.7  \\ 
           13 & 20091213-034209UTC$_{-}$crab-NTT & December 13, 03:51:47.5 & 55178.160966 &  178.3  \\
		   14 & 20091213-035043UTC$_{-}$crab-NTT & December 13, 04:00:33.5 & 55178.167054 & 1798.2  \\
		   15 & 20091213-042211UTC$_{-}$crab-NTT & December 13, 04:31:47.5 & 55178.188744 &    6.1  \\
           16 & 20091214-023608UTC$_{-}$crab-NTT & December 14, 02:46:08.7 & 55179.115379 & 7198.2  \\
           17 & 20091214-043954UTC$_{-}$crab-NTT & December 14, 04:49:32.8 & 55179.201074 & 7198.3  \\
		   18 & 20091215-031945UTC$_{-}$crab-NTT & December 15, 03:29:34.9 & 55180.145543 & 7198.6  \\
		   19 & 20091216-021802UTC$_{-}$crab-NTT & December 16, 02:27:33.9 & 55181.102476 & 3598.5  \\
		   20 & 20091216-031834UTC$_{-}$crab-NTT & December 16, 03:28:05.9 & 55181.144513 &  898.3  \\
 \hline
\end{tabular}
\end{minipage}
\end{table*}

The optical light curve of the Crab pulsar is well known. It is characterized by a double-peak folded profile, separated in phase by $\sim 0.4$ (or $\sim 140^0$). It is very stable \citep{zampieri+11}, although marginal evidence for a secular decrease in luminosity \citep{nasuti+96} and weak variations in the pulse shape \citep{karpov+07} have been reported. 

Absolute optical timing of the Crab pulsar light curve has been addressed in some past investigations. For the other optical pulsars, with the sole exception of PSR B0540-69 \citep{middleditch+87,gradari+11}, the weakness of the targets prevented this type of studies till now. After the early attempts of \citet{nelson+70}, \citet{papaliolios70} and \citet{horowitz+71}, long-term (2 through 10 years) timing solutions for the Crab pulsar in the optical band were obtained by \citet{boynton+72}, \citet{groth75a,groth75c}, and \citet{lohsen81}, who found that the pulsar spin-down has a complex behaviour, including a secular slow down accounted for (on average) by a cubic polynomial. Significant phase noise was observed in form of glitches, jumps and random walks in frequency, but their origin remains as yet unknown. From an investigation of the radio timing of a sample of pulsars (not including the Crab) it has been proposed that often this behaviour may be induced by the abrupt change between two different spin-down rates resulting from variations in the pulsars magnetosphere \citep{lyne+10}.

Recent investigations focused on the delay between the arrival time of the main pulse between the radio and other wavelengths, including the optical (e.g. \citealt{oosterbroek+06,oosterbroek+08}), and showed that the pulses in different energy bands are not aligned. This has profound implications for the geometry of the pulsar emission regions. All these studies have shown the relevance of independent timing studies in the optical (and other energy bands) to understand the pulsar emission mechanism by comparison with the radio. Our observations of the Crab pulsar with Iqueye are aimed at obtaining the most accurate optical timing needed to address these issues.

The plan of the paper is the following. Section~\ref{sec2} presents the Iqueye observations of the Crab pulsar and the adopted data reduction procedure. In Section~\ref{sec3} we show the phase fitting of the 2009 January and December data while, in Section~\ref{sec4}, a comparison with the results from the radio ephemerides is presented. Section~\ref{sec5} is devoted to the phase coherent phase fit of the whole year 2009 data (some mathematical details are reported in Appendix~\ref{app}) and Section~\ref{sec6} to some conclusive remarks.

\section{Observations and data reduction}
\label{sec2}

The Crab nebula pulsar was observed with Iqueye mounted at the ESO 3.5 m New Technology Telescope (NTT)
telescope in two separate runs during 2009 January (from 15 through 20) and December (from 13 through 16).
A log of the 20 observations performed in the two runs is reported in Table~\ref{tab1}.
All of them were done in white light. The average net count rate of the Crab pulsar during
the observations was $\sim 4000$ counts s$^{-1}$.

Iqueye is a very fast optical photon counter based on single photon avalanche 
photodiodes (SPAD) Silicon detectors and equipped 
with a digital acquisition system that can record the photon arrival times of each single
photon, similarly to what is usually done in High Energy and Space Astronomy. The
relative/absolute timing accuracy is $\sim$0.2$-$0.6/1 ns for half-an-hour long observations.
For a detailed description of the optomechanical design, data acquisition and control system, 
and data flow handling of the instrument we refer to \cite{naletto+09}.
After a split on a four-faced pyramid, Iqueye directs the incoming beam of light to four SPAD detectors. For the present analysis, their counts were summed together, so that the instrumental polarization coming from the four faces reflections was effectively averaged out.
We considered also additional polarization-dependent effects possibly introduced by the reflection of the incoming light on the M3 mirror of NTT. However, no significant systematic variations (exceeding the statistical uncertainty) of the 2 s folded pulse profiles used for measuring the phase of the main peak (see Section~\ref{sec3}) were noticed during hours long observations.
We note that, in order to decrease the Crab nebula background and maximize the signal-to-noise
ratio, during the 2009 observing runs Iqueye was operated with an entrance pinhole of $1.5"$
radius, commensurate with the seeing conditions during the observations.

During the first observing night we carefully checked for possible systematic frequencies 
in the data induced by telescope guiding and aperture loss effects. We did it by
computing Fourier power spectra of several test targets. Apart from a low-frequency component
caused by a digitization error of the azimuthal encoder of the telescope and promptly removed 
by the ESO staff, we found no spurious signals in the frequency range 1-2000 Hz.
As a further check, a Fourier analysis of the Crab signal was performed for every 
observation and compared with that expected on the basis of the radio and our own ephemerides.
Finally, we simulated also the system behaviour assuming small periodic intensity variations of the signal caused by telescope guide, wind shaking and similar effects, which could be amplified by the rather tight match between pinhole size and seeing. We then estimated the high frequency noise possibly induced by a sinusoidal intensity noise of small amplitude ($<10$\% of the average pulsar signal) and found that it has negligible effects ($<$ a few $\mu$s) on the determination of the peak arrival time for sampling frequencies below a few Hz (which includes our adopted light curve sampling time of 2s; see Section~\ref{sec3}). However, this additional phase noise may explain the excess width of the distribution of residuals that we noticed in some observations (see Section~\ref{sec5}).

\begin{table}
 \centering
 \begin{minipage}{84mm}
 \caption{Geocentric coordinates of the ESO 3.5 m NTT telescope in La Silla.
      The $3\sigma$ uncertainty is 0.3 m.}
  \label{tab2}
  \begin{tabular}{@{}lll@{}}
  \hline
            $x$ & $y$ & $z$ \\
            (m) & (m) & (m) \\
 \hline
  1838193.8 &  $-5258983.6$  &  $-3100153.5$  \\
 \hline
\end{tabular}
\end{minipage}
\end{table} 

\begin{figure}
 \includegraphics[width=84mm]{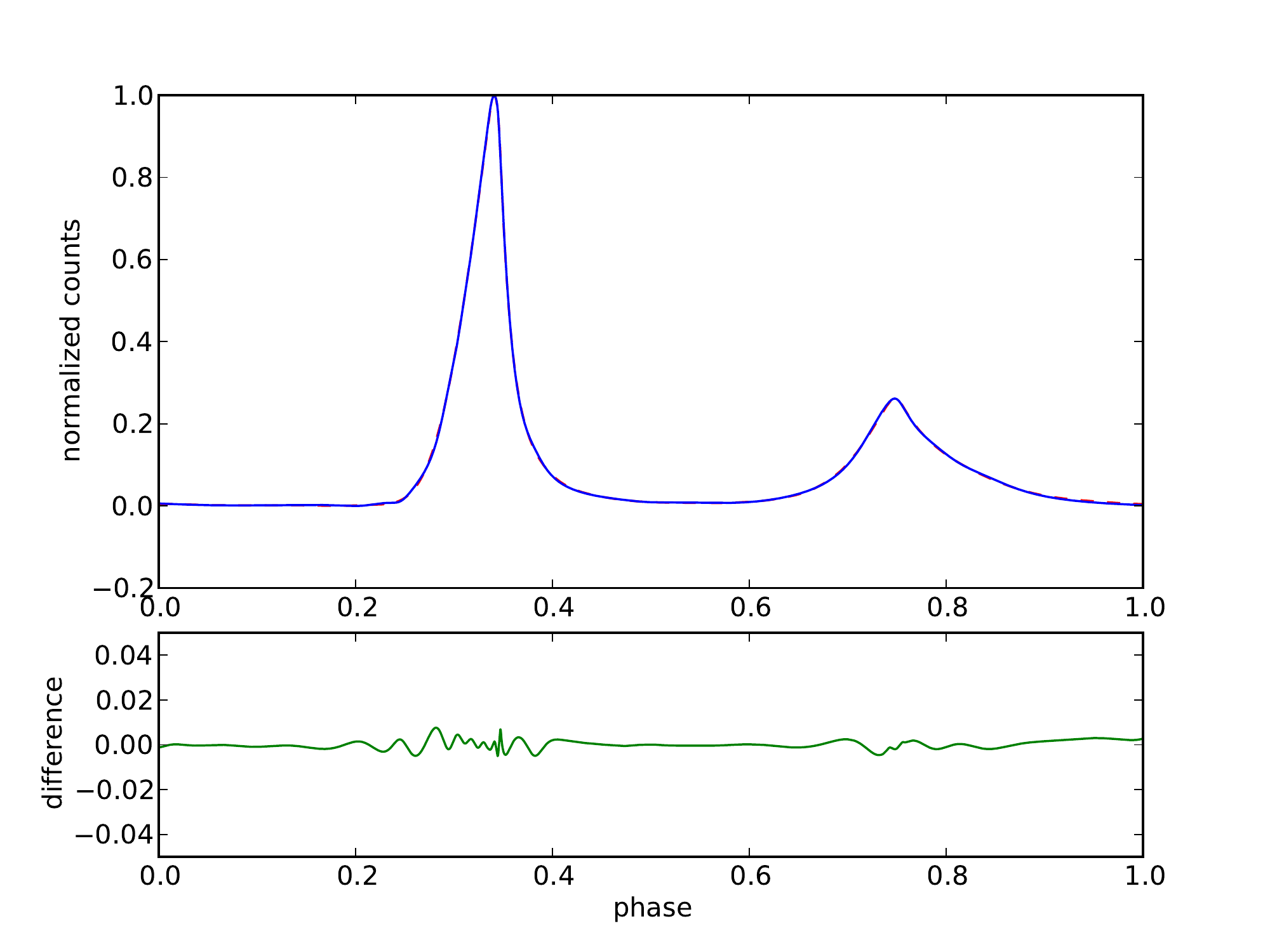}
 \caption{
 {\it Upper} panel: Analytic template (solid line) adopted to fit the Crab pulse shape (method b in the text), obtained from the sum of 16 Lorentzians. The smoothed pulse shape of the whole data set (see text for details) is beneath the solid one.
 The two curves are normalized to the maximum of the smoothed pulse shape. The difference between the two curves is shown in the {\it bottom} panel. The interval adopted to perform the fit is between -0.07 and 0.54 from the position of the main peak. In this interval the difference is at most $\simeq 0.006$ on the tails of the peaks (where the normalized counts $\sim 0.1$), which corresponds to a fractional difference $\simeq 6$\%.
 }
 \label{figtempl}
\end{figure}

\begin{figure*}
 \includegraphics[width=84mm]{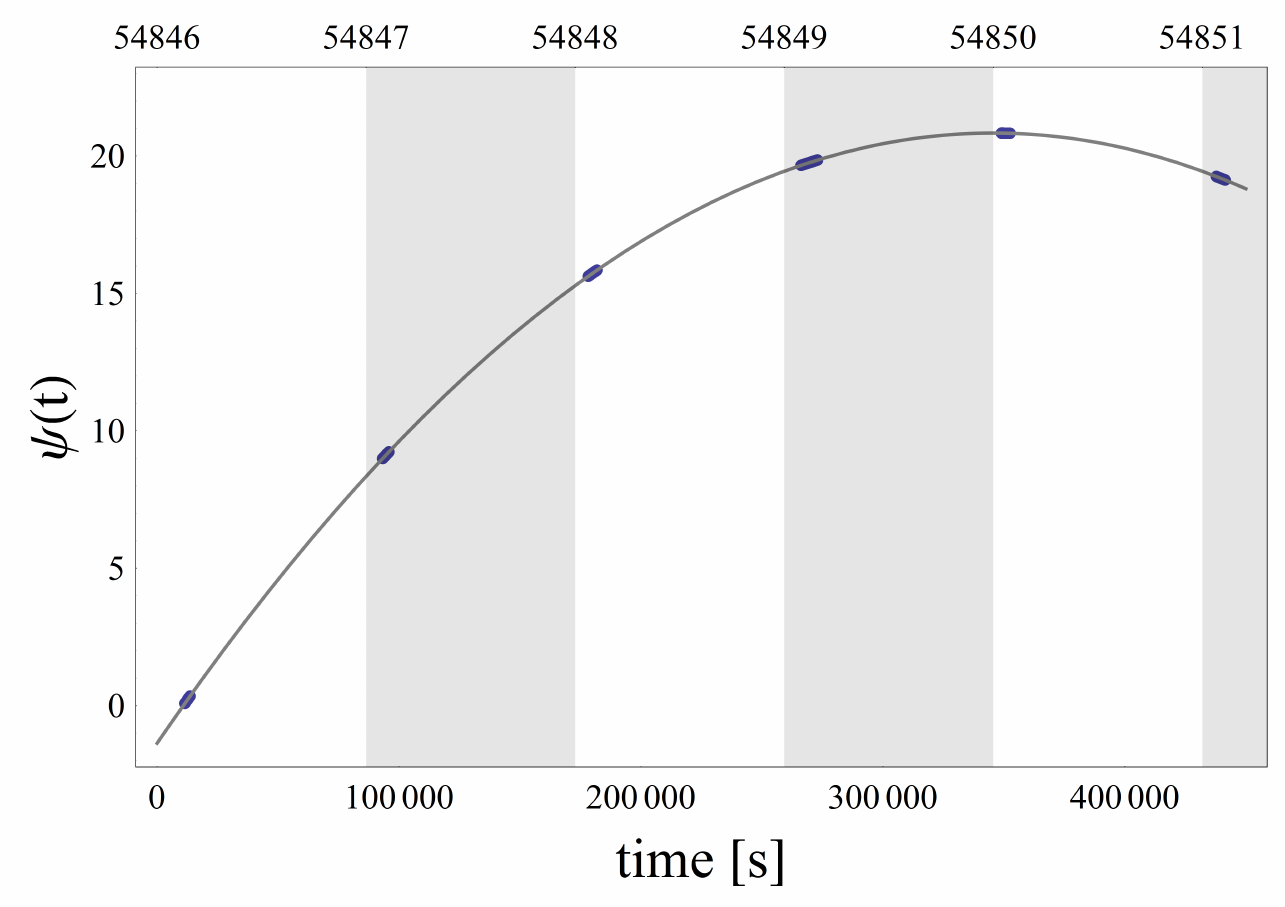}
 \includegraphics[width=84mm]{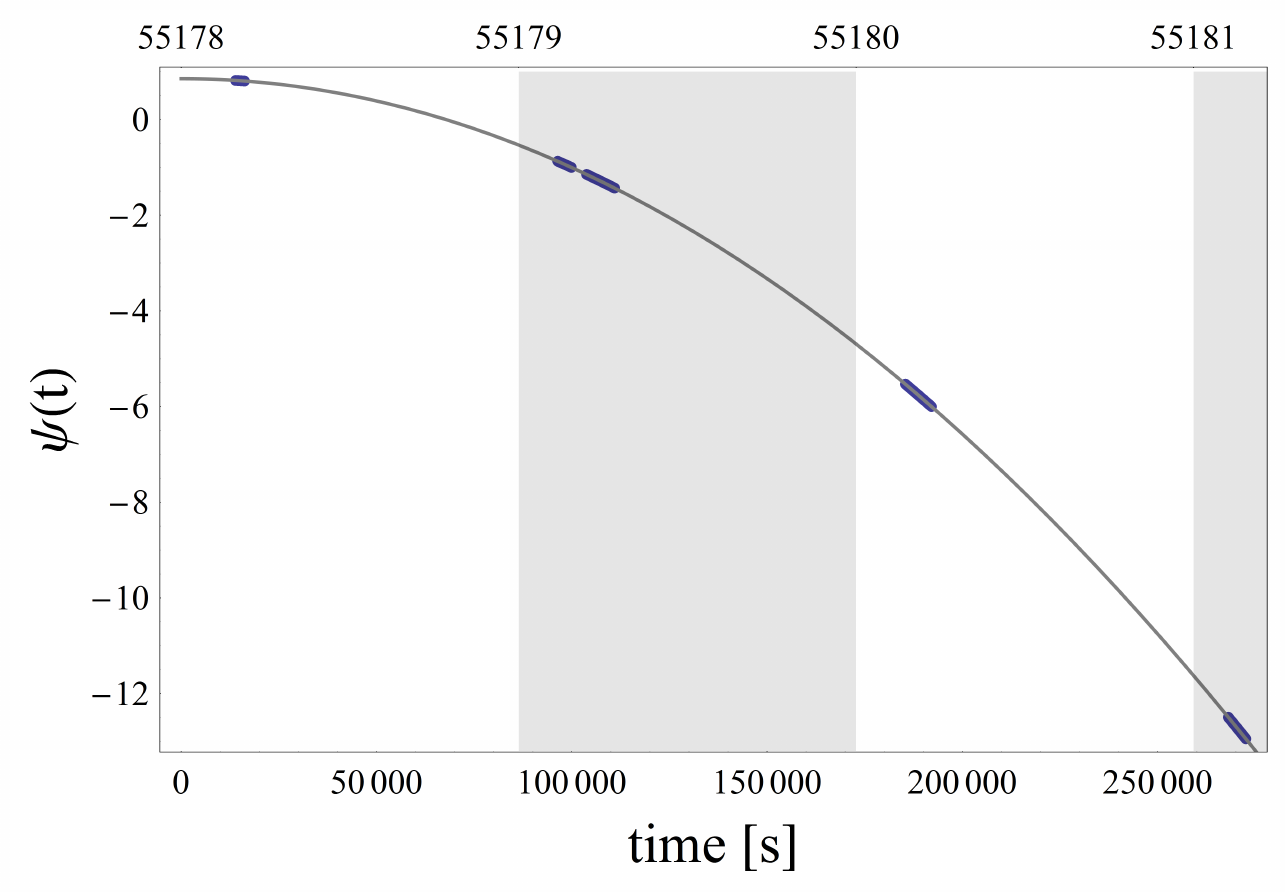}
 \caption{Phase $\psi$ of the Crab pulsar (with respect to uniform rotation) during the January (left) and December (right) Iqueye runs. Time on the top axis is expressed in Modified Julian Date (MJD). The timing solution does not include the cubic term.}
 \label{fig2}
\end{figure*}

The data reduction procedure consists of converting the arrival time tags of each photon in local UTC time 
by means of our own developed software (QUEST v. 1.1.1; P. Zoccarato, internal technical report). Times are then referred to the Solar system
barycentre using the software TEMPO2\footnote{http://www.atnf.csiro.au/research/pulsar/ppta/tempo2}
\citep{hobbs+06,edwards+06}. For the sake of comparison with the Jodrell Bank monthly 
ephemerides \citep{lyne+93}\footnote{http://www.jb.man.ac.uk/$\sim$pulsar/crab.html \label{footjb}}, we decided 
to use TEMPO2 in TEMPO1 emulation mode and to adopt the same position for the Crab pulsar  
(RA 05$^{\rm h}$34$^{\rm m}$31$^{\rm s}\!.97232$, Dec. +22$^0$00$'$52$''\!.0690$ [J2000]), with no correction for proper motion.
To perform barycentric corrections, the software also needs an accurate value of the observatory
geocentric coordinates (Table~\ref{tab2}). They were measured with a GPS receiver 
connected to an antenna situated at the dome of the telescope. 
The final position was referred to the intersection of the telescope azimuth and elevation axes by laser assisted metrology.

\begin{table*}
 \centering
 \begin{minipage}{120mm}
  \caption{Analytic template (16 Lorentzians) adopted to fit the Crab pulse shape ($p$, $q$, $x_1$ free parameters): $f(x) = p \sum_{i=1}^{16} d_{i-1} b_i^2/[b_i^2 + (x-x_1+h_{i-1})^2] + q$}
  \label{tablor}
  \begin{tabular}{@{}llllll@{}}
  \hline
  Parameter & Value & Parameter & Value & Parameter & Value \\
  \hline
  $b_1$     &  0.0146996    &  $d_0$     &         1          &     $h_0$     &  0           \\
  $b_2$     &  0.0146996    &  $d_1$     &         0.217538   &     $h_1$     &  0.0295389   \\
  $b_3$     &  0.0146996    &  $d_2$     &         0.120438   &     $h_2$     &  0.0452724   \\
  $b_4$     &  0.0146996    &  $d_3$     &         0.343795   &     $h_3$     &  0.0159706   \\
  $b_5$     &  0.0146996    &  $d_4$     &         0.0274555  &     $h_4$     &  -0.0405742  \\
  $b_6$     &  0.00390605   &  $d_5$     &         0.104503   &     $h_5$     &  -0.004064   \\
  $b_7$     &  0.0131649    &  $d_6$     &         0.0524991  &     $h_6$     &  -0.408426   \\
  $b_8$     &  0.0517911    &  $d_7$     &         0.0462601  &     $h_7$     &  -0.493455   \\
  $b_9$     &  0.0386609    &  $d_8$     &         0.250336   &     $h_8$     &  -0.400741   \\
  $b_{10}$  &  0.0377745    &  $d_9$     &         0.063293   &     $h_9$     &  -0.445372   \\
  $b_{11}$  &  0.0156592    &  $d_{10}$  &        -0.0323015  &     $h_{10}$  &  0.0948912   \\
  $b_{12}$  &  0.0325165    &  $d_{11}$  &        -0.0176647  &     $h_{11}$  &  0.133417    \\
  $b_{13}$  &  0.0531056    &  $d_{12}$  &        0.0128576   &     $h_{12}$  &  0.355586    \\
  $b_{14}$  &  0.209385     &  $d_{13}$  &        0.00944315  &     $h_{13}$  &  -0.0200141  \\
  $b_{15}$  &  -0.0630249   &  $d_{14}$  &        -0.00883256 &     $h_{14}$  &  -0.261205   \\
  $b_{16}$  &  0.0259154    &  $d_{15}$  &        -0.00388652 &     $h_{15}$  &  -0.153419   \\
  \hline
  \end{tabular}

\end{minipage}
\end{table*}

\section{Phase fitting 2009 January and December runs}
\label{sec3}

Each barycentric corrected time series is divided in short segments 2 s long
and folded over a reference period $P_{\rm init}$.
In order to cross check the accuracy of our results, in each segment the phase of the main 
pulse is determined using two different methods: (a) calculating the maximum of the cross-correlation function between the 2 s folded profile and a smoothed pulse shape of the 
whole data set (see \citealt{germana+12}). The latter is a numerical template, obtained by smoothing the distribution generated by the counts of all observations distributed according to their phase of arrival; 
(b) fitting the main peak and the interpulse in the 2-s folded profile with an analytic template, 
constructed to provide an accurate representation of the smoothed pulse shape of the 
whole data set (sum of Lorentzians; see Fig.~\ref{figtempl} and~Table~\ref{tablor}). 
The fitting parameters are the total amplitude $p$, 
the background level $q$, and the centroid of the first Lorentzian $x_1$\footnote{The peak of the analytic template is at $x_{max} =  x_1 - 9.6 \times 10^{-5}$. However, the phase of the main pulse in each segment is taken to be $x_p =  x_1 + 1.2 \times 10^{-5}$, where the shift ($1.2 \times 10^{-5}$) is inferred from fitting the smoothed pulse shape with the analytic template itself.}.
The adopted phase bin is $1/300$ in method (a) and $1/1000$ in method (b).
The two algorithms return phase measurements that are in agreement within the errors. The typical spread in phase measurements for 2-s long segments is $\sim 0.005$ in phase, or $\sim 17-18 \, \mu$s in time.

Phase measurements of the main peak track pulsar spin-down with respect 
to uniform rotation. If $\phi$ is the phase of the pulsar main peak and 
$\phi^{'}=\nu_{\rm init}(t-t_{0})$ the phase of a uniform rotator with frequency
$\nu_{\rm init}=1/P_{\rm init}$, the actual measured phase is
\begin{equation}
\psi(t)=\phi(t)-\phi^{'}(t)=\phi(t)-\frac{1}{P_{\rm init}}(t-t_0) \, ,
\label{psi}
\end{equation}
where $t_0$ is the time (the starting Julian date of the observing run) at which $\phi$ and $\psi$ agree up to an integer number of $\phi$ turns. $\psi$ represents the difference between the phase of the pulsar and that of the uniform rotator and its fractional part is determined by cross-correlating or fitting the measured pulse shape as described above.
The integer part is found as explained below.

Since the pulsar frequency changes only slowly in time, the phase $\psi$ is modelled with a timing solution in the form of a second-order polynomial:
\begin{equation}
\psi(t)=\phi_{0}+(\nu_0-\nu_{\rm init})(t-t_{0})+\frac{1}{2}\dot{\nu}_0(t-t_{0})^{2} \, ,
\label{eq1}
\end{equation}
where $\phi_0$, $\nu_0$, $\dot{\nu}_0$ are the actual pulsar phase, rotational frequency and its first derivative at $t_0$, respectively. The reference time is $t_0^{\rm J}=54846$ MJD for January and $t_0^{\rm D}=55178$ MJD for December. On intervals of a few days, this expression is statistically adequate to describe the phase behaviour. On longer time intervals, higher order terms become important (see Section~\ref{sec5}).

\begin{table}
 \centering
 \begin{minipage}{80mm}
  \caption{Timing solutions for the two separate 2009 January and December data sets.}
  \label{tab3}
  \begin{tabular}{@{}lll@{}}
  \hline
     &      2009 January observations$^a$ \\
	 \hline 
  $t_0$ (MJD)                 &  54846.0                                                 \\
  $\phi_0$                    &  $0.632498 \pm 5.0 \times 10^{-5}$                       \\
  $\nu_0$ (s$^{-1}$)          &  $29.73966616327 \pm 4.4 \times 10^{-10}$                \\
  $\dot{\nu}_0$ (s$^{-2}$)    &  $-3.717790 \times 10^{-10} \pm 1.7 \times 10^{-15}$     \\
  $P_0$ (s)                   &  $0.0336251252623 \pm 5.0 \times 10^{-13}$               \\
  $t_{\rm arr}$              (s)$^c$  &  $0.0212678 \pm 1.7 \times 10^{-6}$                     \\
  \hline
     &      2009 December observations$^b$ \\
	 \hline
  $t_0$ (MJD)                 &  55178.0                                                  \\
  $\phi_0$                    &  $0.853410 \pm 3.6 \times 10^{-5}$                        \\
  $\nu_0$ (s$^{-1}$)          &  $29.72900668822 \pm 5.2 \times 10^{-10}$                 \\
  $\dot{\nu}_0$ (s$^{-2}$)    &  $-3.714336 \times 10^{-10} \pm 3.2 \times 10^{-15}$      \\
  $P_0$ (s)                   &  $0.0336371817090 \pm 5.8 \times 10^{-13}$                \\
  $t_{\rm arr}$              (s)$^c$  &  $0.0287063 \pm 1.2 \times 10^{-6}$                       \\
  \hline  
\end{tabular}

$^a$ $P_{\rm init}=0.0336252705350$ s \par
$^b$ $P_{\rm init}=0.0336371817073$ s \par
$^c$ Arrival time of the main peak after $t_0$: $t_{\rm arr}=\phi_0 P_{\rm init}$ \par
\end{minipage}
\end{table}

Fig.~\ref{fig2} shows the timing solutions for January and December 2009 obtained fitting equation~(\ref{eq1}) to the measured phases by means of a $\chi^2$ minimization procedure. 
The adopted reference periods $P_{\rm init}$ are 0.0336252705350 s and 0.0336371817073 s, respectively, and the reference epochs are those given in Table~\ref{tab3}. 
Since the correlation/fitting procedure can only give the fractional part of the phase, the integer part of the phase difference between successive nights are found by minimizing $\chi^2$ with respect to integer variations of $\psi$.
The best$-$fitting parameters are reported in Table~\ref{tab3}, along with the arrival time of the main peak after midnight $t_{\rm arr}=\phi_0 P_{\rm init}$ and the pulsar rotational period $P_0=1/\nu_0$ at $t_0$. Quoted errors are purely statistical and correspond to the $2\sigma$ confidence level for one interesting parameter.

\section{Comparison with Jodrell Bank radio ephemerides: optical$-$radio phases}
\label{sec4}

We can compare rotational phases and arrival times of the main peak in the optical with phases and arrival times in the radio, reported in the Jodrell Bank Crab Pulsar Monthly Ephemeris website (JB ephemerides hereafter; \citealt{lyne+93}, see footnote~\ref{footjb}).
Radio phases are calculated using equation~(\ref{eq1}), where the values of $\phi_{0,r}$, $\nu_{0,r}$, $\dot{\nu}_{0,r}$ are computed from the closest available epoch reported in the JB archive (usually day 15 of each month).

\begin{figure*}
 \includegraphics[width=84mm]{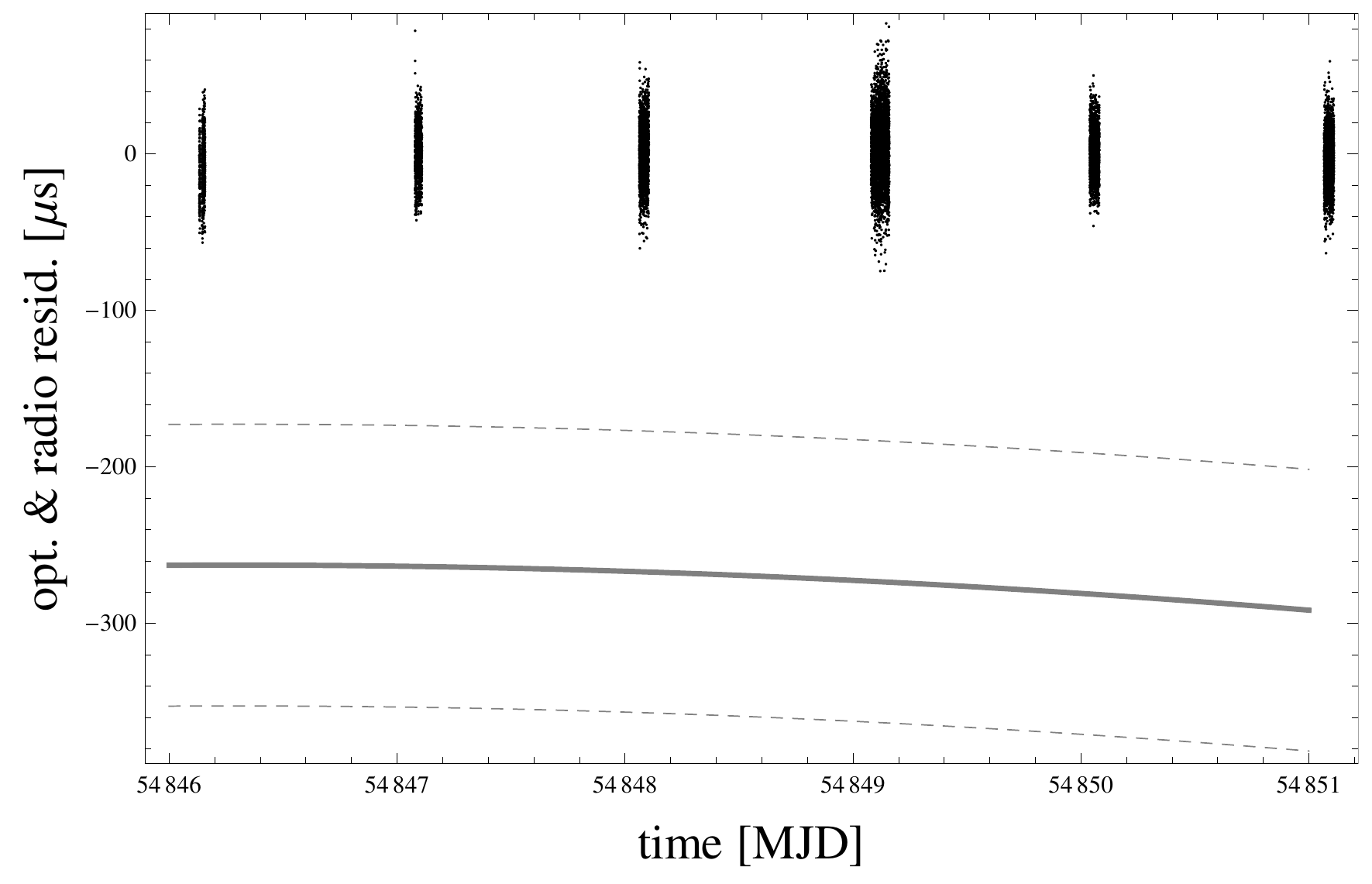}
 \includegraphics[width=84mm]{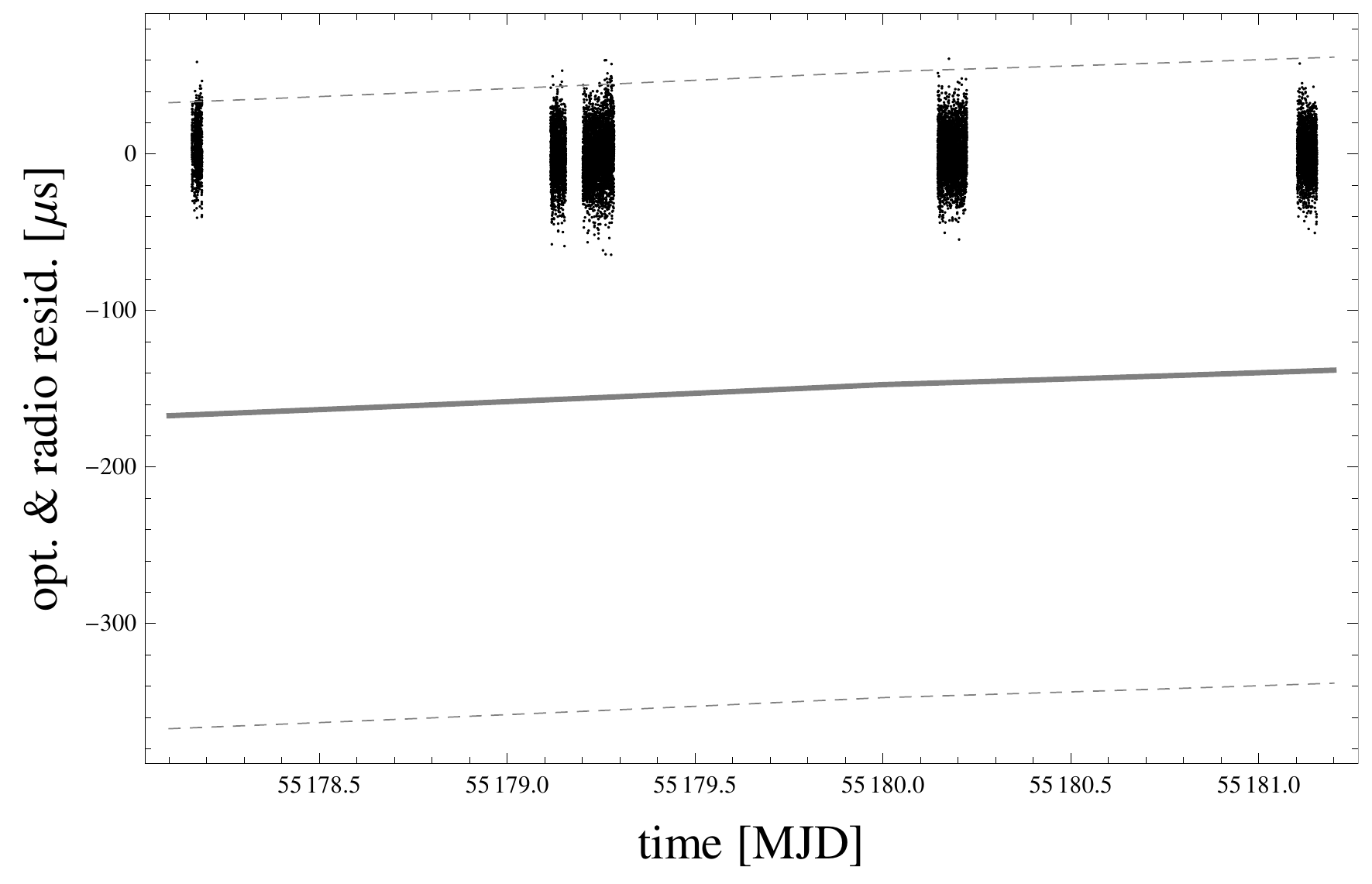}
 \caption{Phase residuals for the January (left) and December (right) Iqueye runs.
 The solid line represents the difference between the optical and radio timing solutions, while the dashed lines bound the total (statistical plus systematic) uncertainty on the radio phases. The residuals and optical$-$radio phase difference are computed from the global fit of all the 2009 Iqueye observations (see Section~\ref{sec5}).
 }
 \label{fig3}
\end{figure*}

Fig.~\ref{fig3} shows the difference between the optical and radio timing solutions. The residuals and optical$-$radio phase difference are computed from the global fit of all the 2009 Iqueye observations (see Section~\ref{sec5}). Considering only the statistical error on the optical data, the average difference between the time of arrival of the optical and radio peaks is $\sim 240 \, \mu$s in January and $\sim 160 \, \mu$s in December 2009. The total uncertainties (statistical plus systematics) on JB radio ephemerides is 90 $\mu$s in January and 200 $\mu$s in December. Therefore, the optical$-$radio difference is significant only for January data, with the optical peak leading the radio one. 
An independent measurement of the December 2009 optical$-$radio phase shift ($\sim 178 \, \mu$s), based on the same  Iqueye data, was obtained by \citet{collins+11} using simultaneous JB radio observations and is consistent with that reported here.

In addition to the optical$-$radio phase difference, Fig.~\ref{fig3} also indicates the existence of a phase drift. Considering the optical$-$radio phase difference from the fit of each observing run separately, they amount to 3.7$\pm$0.1 and 17.5$\pm$0.2 $\mu$s/d$^{-1}$ for January and December, respectively.

\section{Phase-coherent optical timing solution in 2009}
\label{sec5}

The quality of the Iqueye data is such that we can attempt to connect the phases measured several months apart in the optical band, if we assume that the pulsar is slowing down quite regularly according to a simple braking law.
We recomputed the phases for all the 2009 observations reported in Table~\ref{tab1} following the approach outlined in Section~\ref{sec3} and taking as common reference period for the two runs $P_{\rm init}=0.03362513000000$ s.
For the sake of comparison with the ``local'' January and December timing solutions, two different reference epochs are adopted, MJD 54846 and 55178 (see below).

On a 1 yr time interval the parabolic spin-down model of equation~(\ref{eq1}) is no longer adequate to describe the behaviour of the phases. We then adopted a cubic polynomial of the form (e.g. \citealt{boynton+72}): 
\begin{equation}
\psi(t)=\phi_{0}+(\nu_0-\nu_{\rm init})(t-t_{0})+\frac{1}{2}\dot{\nu}_0(t-t_{0})^{2}+\frac{1}{6}\ddot{\nu}_0(t-t_{0})^{3}
\label{eq2}
\end{equation}
where $\ddot{\nu}_0$ is the second derivative of the rotational frequency at $t_0$. In several spin-down models the pulsar frequency derivative $\dot \nu$ is predicted to satisfy the well-known braking index relation:
\begin{equation}
\dot \nu = - K \nu^n \, ,
\label{brind}
\end{equation}
where $K$ is a constant and $n$, the braking index, depends on the physical mechanism responsible for the loss of angular momentum and rotational energy of the star. 
Using this expression and adding the fourth-order term, equation~(\ref{eq2}) can be cast in the form
\begin{eqnarray}
\psi (t) &=& \phi_0 + (\nu_0-\nu_{\rm init})(t-t_{0}) + \frac{1}{2} {\dot \nu_0}(t-t_0)^2  \nonumber \\
&+& \frac{n}{6} \frac{\dot \nu_0^2}{\nu_0}(t-t_0)^3  \nonumber \\
&+& \frac{n(2n-1)}{24}\frac{\dot \nu_0^3}{\nu_0^2}(t-t_0)^4  \, .
\label{eqbrind}
\end{eqnarray}
We can estimate the importance of the parabolic, cubic and fourth-order terms in the Taylor expansion~(\ref{eqbrind}). The spin-down is sensitive to them if they contribute at least a significant number of rotations, say $\Delta \psi \approx 10$. Assuming $\nu_0 \approx 30$ Hz and $\dot \nu_0 \approx - 4\times 10^{-10}$ s$^{-2}$, this requires a time $\Delta t_2 = |2\Delta \psi/{\dot \nu_0}|^{1/2} \approx 2.5$ d for the parabolic term, $\Delta t_3 = (6\Delta \psi \nu_0/(n {\dot \nu_0^2}))^{1/3} \approx 180$ d for the cubic term, and $\Delta t_4 = |24\Delta \psi \nu_0^2/{((2n^2-n) \dot \nu_0^3)}|^{1/4} \approx 4$ yr for the fourth-order term to become important. Therefore, while it is meaningful to adopt equation~(\ref{eq1}) for a single observing run, a full timing solution over 1 yr requires a cubic spin-down representation as in equation~(\ref{eq2}) (e.g. \citealt{boynton+72}). Higher order terms are needed for longer time intervals.
 

The phase fit with the cubic spin-down model of equation~(\ref{eq2}) of all the 2009 (January and December) optical phase measurements has been performed in two different ways. In the first approach, the pulsar phases  determined in Section~\ref{sec3} ($\psi$ phases) are directly fitted adding a phase shift to the December data set as described below. In the second approach, the true pulsar rotational phases ($\phi$ phases) are reconstructed for each observing run and the fit is performed after determining the integer phase shift between January and December, as explained below. The two approaches return results in agreement within 3$\sigma$. They are shown in Table~\ref{tab4}. 
For the sake of comparison with the ``local" January and December timing solutions, in the first approach two independent fits of the whole data set are performed, assuming as reference epochs MJD 54846 or 55178, respectively. The quoted errors are purely statistical and correspond to the $2\sigma$ confidence level for one interesting parameter.

To perform the fit using the first approach, the two data sets of January and December had to be referred to a common reference period $\hat P_{\rm init}$. Phase folding becomes inaccurate if there is a significant mismatch between $\hat P_{\rm init}$ and the actual pulsar period, which is different in January and December. Therefore, phases were first computed using the reference periods reported in Table~\ref{tab3} and then referred to $\hat P_{\rm init}$ for both the January and December data using the two expressions (see Appendix~\ref{app}):
\begin{equation}
\hat \psi_{\rm J}(t) = u \, f\!r[\psi_{\rm J}(t)] - \left( u - 1 \right) n(t) - v - {\hat \psi_{n,{\rm J}}}
\label{eqpsij}
\end{equation}
\begin{equation}
\hat \psi_{\rm D}(t) = u \, f\!r[\psi_{\rm D}(t)] - \left( u - 1 \right) n(t) - v - {\hat \psi_{n,{\rm D}}} - \Delta \hat \psi_n \, ,
\label{eqpsid}
\end{equation}
where $u=P^{{\rm J},{\rm D}}_{\rm init}/\hat P_{\rm init}$, $n(t)={\rm int}[(t_{\rm P}-t_0^{{\rm J},{\rm D}})/P_{\rm init}^{{\rm J},{\rm D}}]={\rm int}[(t-t_0^{{\rm J},{\rm D}})/P_{\rm init}^{{\rm J},{\rm D}}]$\footnote{$t_{\rm P}(t)$ is the time of arrival of the main pulse after time $t$.}, $v=f\!r[(t_0^{{\rm J},{\rm D}}-\hat t_0)/\hat P_{\rm init}]$ and $f\!r[ \, ]$ denotes the fractional part. 
Equations~(\ref{eqpsij}) and~(\ref{eqpsid}) come from the requirement that the time of arrival of a single pulse determined in the two different ``temporal gauges'' be equal (or, similarly, that the phase $\phi$ of the main pulse computed using $\psi$ or $\hat \psi$ be equal; see Appendix~\ref{app}). 
We adopted $\hat P_{\rm init}=0.03362513000000$ s (Table~\ref{tab4}).
The phase shifts $\hat \psi_{n,{\rm J}}$ and $\hat \psi_{n,{\rm D}}$ are the integer part of the phase difference between successive nights in January and December, respectively, and are found by minimizing $\chi^2$ with respect to integer variations of $\hat \psi$, as explained in Section~\ref{sec3}. An additional integer shift $\Delta \hat \psi_n$ is needed to phase connect together the January and December data sets and is also determined through $\chi^2$ minimization.

In the second approach, we express the phase $\phi$ using equation~(\ref{psi}) for both the January and December data as
\begin{eqnarray}
&& \phi_{\rm J}(t)=\psi_{\rm J}(t)+\frac{1}{P^{\rm J}_{\rm init}}(t-t_0^{\rm J}) \label{phijandec1} \\
&& \phi_{\rm D}(t)=\psi_{\rm D}(t)+\frac{1}{P^{\rm D}_{\rm init}}(t-t_0^{\rm D})+N \, ,
\label{phijandec2}
\end{eqnarray}
where $N$ is an integer phase shift. The union of the two sets of data is fitted to a cubic polynomial with the unknown integer $N$ as a fitting parameter, which is varied to produce the minimum value of $\chi^2$.

Up to third order, the braking index model (equation~\ref{eqbrind}) is essentially indistinguishable from the cubic spin-down model, both having four free parameters. In fact, the pulsar braking index can be determined directly fitting  equation~(\ref{eqbrind}), truncated to third order, to the data or from the pulsar frequency and its first and second derivatives: $n=-\nu_0 \ddot{\nu}_0/\dot{\nu}^2_0$. The measured value is again reported in Table~\ref{tab4}. 

\begin{table}
 \centering
 \begin{minipage}{80mm}
  \caption{Timing solutions for whole year 2009. Two solutions are reported, corresponding to two different reference epochs (MJD 54846 and 55178, respectively.}
  \label{tab4}
  \begin{tabular}{@{}lll@{}}
  \hline
                              &   All 2009 Iqueye observations$^a$  \\
  \hline
  $\hat t_0$ (MJD)             &   54846                                                   \\
  $\phi_0$                    &   $0.632777 \pm 3.2\times 10^{-5}$                        \\
  $\nu_0$ (s$^{-1}$)          &   $29.73966616036 \pm 1.1\times 10^{-10}$                 \\
  $\dot{\nu}_0$ (s$^{-2}$)    &   $-3.71769275 \times 10^{-10} \pm 1.8\times 10^{-17}$    \\
  $\ddot{\nu}_0$ (s$^{-3}$)   &   $1.13183 \times 10^{-20} \pm 1.2 \times 10^{-24}$       \\
  $P_0$ (s)                   &   $0.03362512526562 \pm 1.2 \times 10^{-13}$              \\
  $t_{\rm arr}$ (s)$^b$               &   $0.0212772 \pm 1.1 \times 10^{-6}$                      \\
  $n^c$                         &   $2.43539 \pm 0.00026$                                   \\
  \hline
  $\hat t_0$ (MJD)             &   55178                                                   \\
  $\phi_0$                    &   $0.853283 \pm 2.2\times 10^{-5}$                        \\
  $\nu_0$ (s$^{-1}$)          &   $29.72900668951 \pm 1.3\times 10^{-10}$                 \\
  $\dot{\nu}_0$ (s$^{-2}$)    &   $-3.71444613 \times 10^{-10} \pm 1.9\times 10^{-17}$    \\
  $\ddot{\nu}_0$ (s$^{-3}$)   &   $1.13183 \times 10^{-20} \pm 1.2\times 10^{-24}$        \\
  $P_0$ (s)                   &   $0.03363718170755 \pm 1.4 \times 10^{-13}$              \\
  $t_{\rm arr}$ (s)$^b$               &   $0.02869175 \pm 7.4 \times 10^{-7}$                     \\
  $n^c$                         &   $2.43878 \pm 0.00026$                                   \\
  \hline
  \end{tabular}

$^a$ $\hat P_{\rm init}=0.03362513000000$ s, $\Delta \hat \psi_n=152784$, $N=852923471$ \par
$^b$ Arrival time of the main peak after $t_0$: $t_{\rm arr}=\phi_0 P_{\rm init}$ \par
$^c$ Braking index: $n=-\nu_0 \ddot{\nu}_0/\dot{\nu}^2_0$
\end{minipage}
\end{table}

\begin{figure}
 \includegraphics[width=84mm]{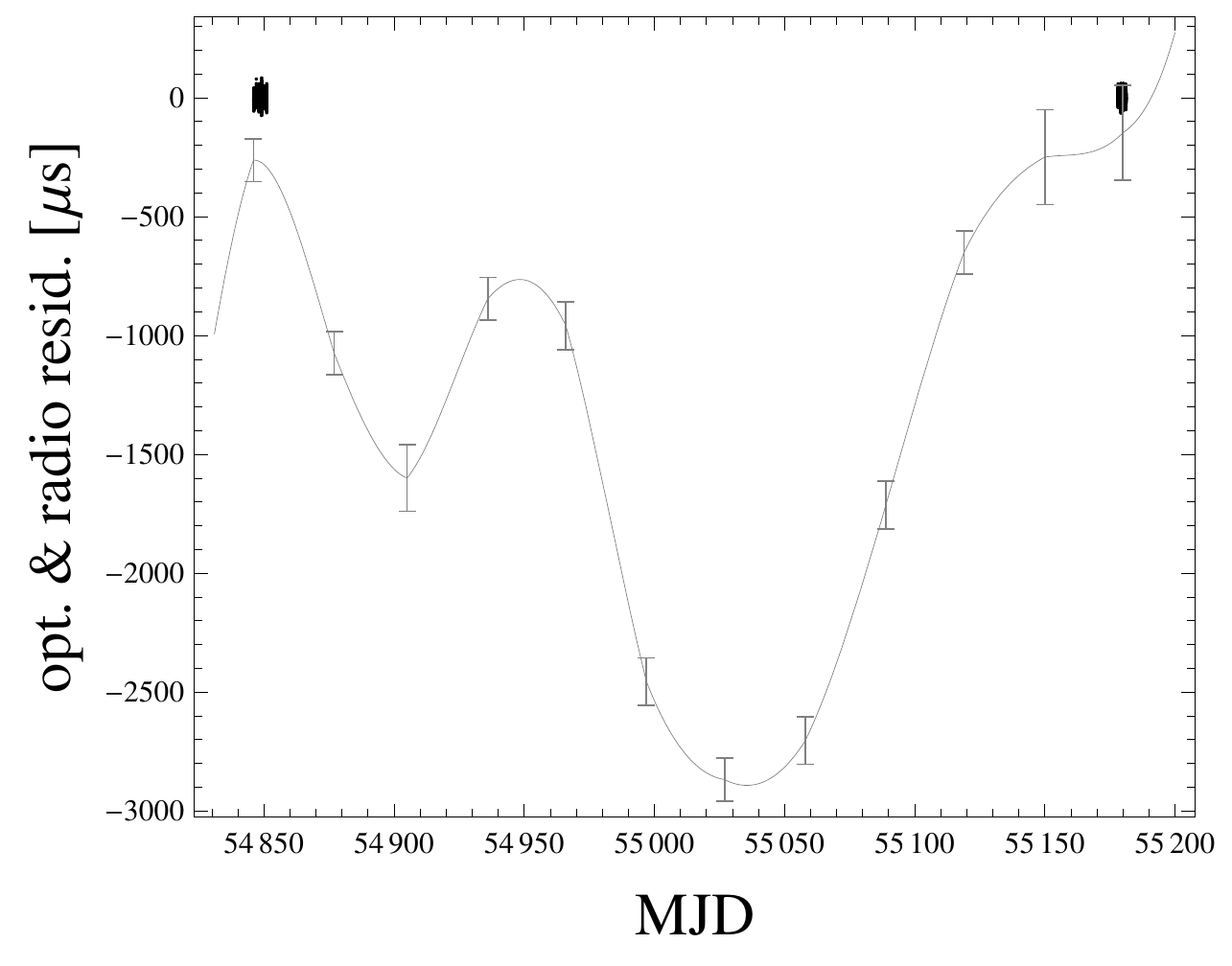}
 \caption{
 Phase residuals for all the 2009 Iqueye observations (black dots) and the JB ephemerides (grey vertical bars) with respect to the optical fit with the cubic spin-down model reported in Table~\ref{tab4} ($\Delta \hat \psi_n=152784$, $N=852923471$). 
 The gray solid line represents a smooth curve joining the average monthly cubic fits of the JB ephemerides. The vertical error bars are the total (statistical plus systematic) uncertainty on the radio phases as quoted in JB ephemerides.}
 \label{fig4}
\end{figure}

A phase coherent timing solution for year 2009 can also be derived from the JB radio ephemerides. The calculation is performed in a similar way, fitting equation~(\ref{eq2}) to the phases reported in the JB archive. The values of radio phase, frequency and its derivatives are usually given at day 15 of each month. They are computed as averages of different measurements performed during the month. From $\phi_{0,r}$, $\nu_{0,r}$, $\dot{\nu}_{0,r}$ at day 15 we extrapolate the phases from $\sim 15$ d before through $\sim 15$ d after the epoch reported in the ephemerides and match them at the beginning of each month. This is done using equation~(\ref{eq2}) and approximating $\ddot{\nu}_{0,r} \approx 2 \nu_{0,r}^3 \dot{\nu}^2_{0,r}$, as suggested in the explanatory set of notes of the JB ephemerides (see footnote~\ref{footjb}).
The matching precision at monthly intervals is consistent with the uncertainty reported in the JB archive ($\sim 100$$-$$200 \, \mu$s). A single cubic global fit of the radio phases is then performed with equation~(\ref{eq2}). 
The $\chi^2$ of the fit is 345 for 12 degrees of freedom, which tells that the assumption of a constant braking law is not valid. Excursions of the radio phases about the cubic fit from the optical data, shown in Fig.~\ref{fig4}, are larger than measurement uncertainties.

The short-term (monthly) radio phase behaviour differs from the one predicted by equations~(\ref{eq2}) or~(\ref{eqbrind}) because of the effects of pulsar rotational variations and irregularities that build up on the radio monthly time-scale, but are not visible on the coarser year-time-scale of our optical observations. However, on average the cubic fit appears to account quite nicely for the long-term phase evolution of the pulsar rotation (see e.g. \citealt{groth75c}).
With the choice of $\Delta \hat \psi_n$ and $N$ described below, the residual timing difference between the radio and optical cubic fits is only a few ms (with small differences also in the frequency, its derivatives, and braking indices).
This deviation is at most $\sim 0.1$ turns, which amounts to $\sim 10^{-10}$ parts of the phase increment on the time interval sampled by observations. We conclude that the pulsar phase follows the constant braking law quite closely, yet not absolutely precisely.

We also note that the braking index inferred from the radio data differs from that obtained in previous epochs. Taking as reference MJD 55000, close to the middle of the interval, $n_{radio}=2.4585 \pm 0.0022$. This is smaller than the average value measured in the radio during years 1969 through 1993 ($n_{radio}=2.51 \pm 0.01$; \citealt{lyne+88,lyne+93}), indicating a change in the strength of the spin-down.

Using the two approaches described above, the integer phase shifts needed to phase connect the December to the January Iqueye data are determined through $\chi^2$ minimization. Four adjacent values of $N$ give a $\chi^2$ which lies in the interval of 90\% statistical significance of the $\chi^2$ distribution for 22525 independent variables (the number of phases measured). They are: 852923566, 852923567 (absolute minimum), 852923568, 852923569. We consider all these four fits acceptable. 
Similarly, four values of $\Delta \hat \psi_n$ lie in the 90\% interval of the $\chi^2$ distribution of the fit: 152784, 152785, 152786 (absolute minimum), 152787.
As discussed above, because of our limited temporal sampling, we were not in a position to detect monthly phase oscillations that the pulsar made during 2009. Therefore, just from optical data we could not decide between the four different values of $N$ or $\Delta {\hat \psi}_n$. Tracking the long-term behaviour with a finer temporal coverage, the radio data allow us to single out the values of $\Delta \hat \psi_n$ (152784) and $N$ (852923469) which have been used to construct the optical fit shown in Fig.~\ref{fig4}.
Using these values, the number of pulsar turns between MJD 54846 and 55178 is in agreement with the value calculated from the JB ephemerides. We obtain $N_{\phi}={\rm int}[\phi_{\rm D}(t_0^{\rm D})-\phi_{\rm J}(t_0^{\rm J})]=852923571$ from equations~(\ref{phijandec1}) and~(\ref{phijandec2}), and $N_{\hat \psi}={\rm int}[{\hat \psi}(t_0^{\rm D})-{\hat \psi}(t_0^{\rm J})+(t_0^{\rm D}-t_0^{\rm J})/{\hat P_{\rm init}}]=852923571$ from equations~(\ref{psi}), (\ref{eqpsij}) and (\ref{eqpsid}).

\begin{figure*}
 \includegraphics[width=154mm]{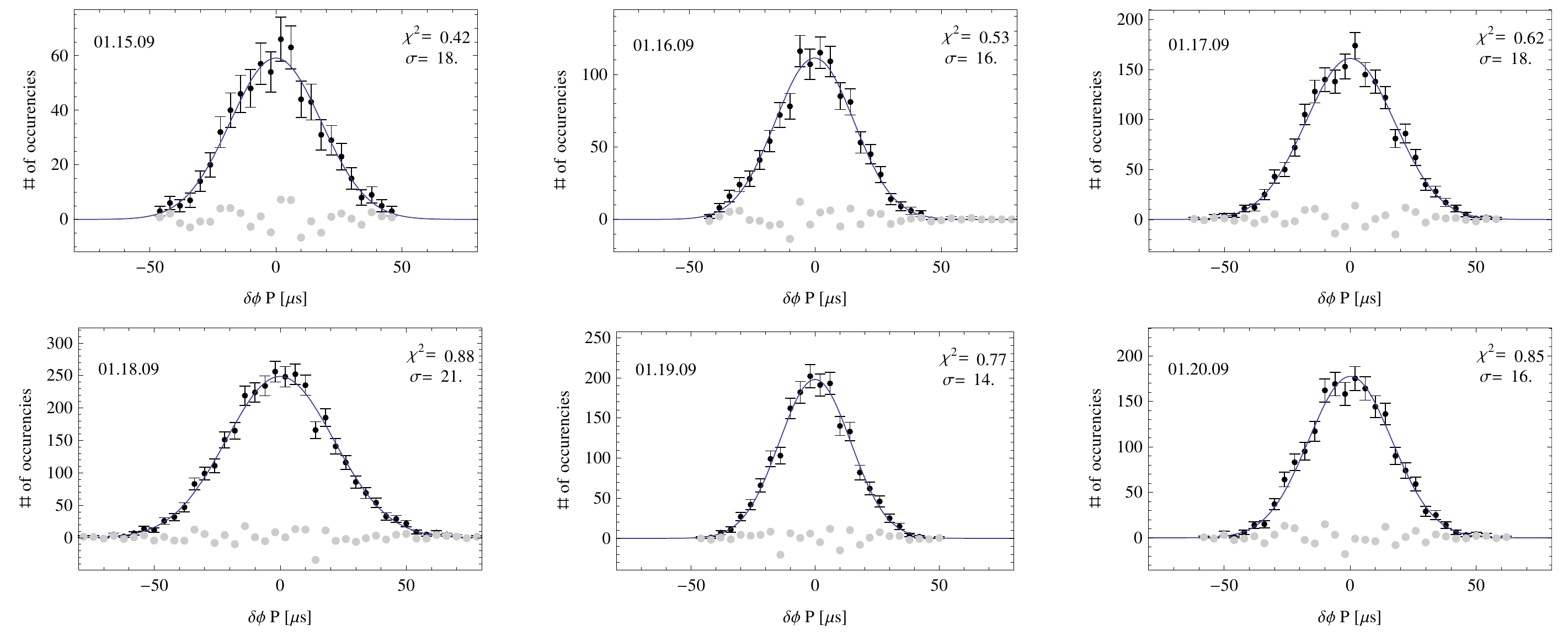}
 \includegraphics[width=124mm]{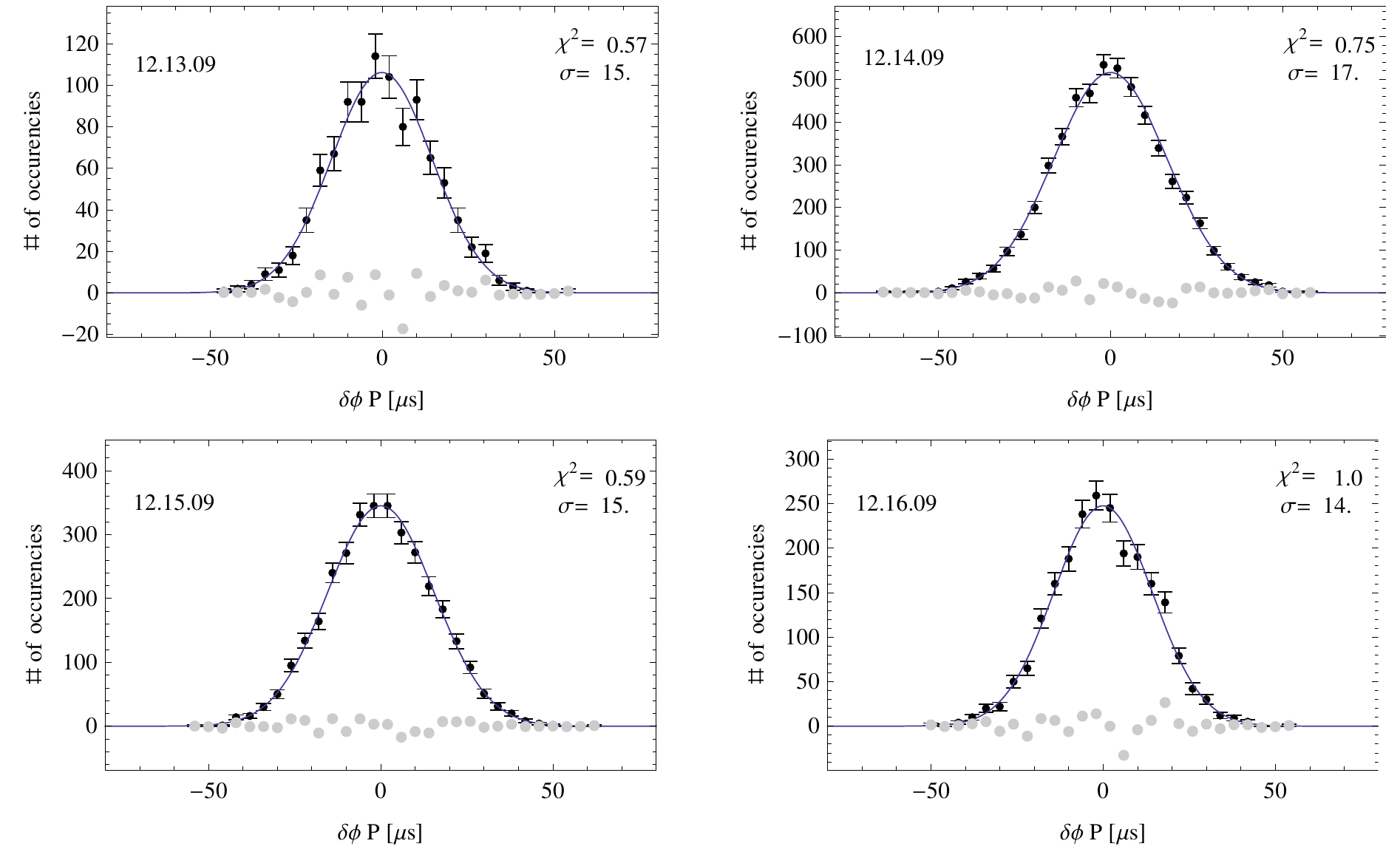} 
 \caption{Distribution of phase residuals for the various 2009 Iqueye observing nights. Data are binned with 32 bins of 2.5$\mu$s each. The solid line is the Gaussian fit, while the grey dots without errorbars represent the difference between the data and the fit. The reduced $\chi^2$ of the fit and the dispersion $\sigma$ of each Gaussian (in $\mu$s) are also shown. From {\it top-left} to {\it bottom-right}: January 15, 16, 17, 18, 19, 20; December 13, 14, 15, 16.}
 \label{fig5}
\end{figure*}



We note that the uncertainties of the parameters quoted in Table~\ref{tab4} are strictly formal statistical 2$\sigma$ confidence intervals, calculated with the assumption that a single $N$ or $\Delta {\hat\psi}_n$ (the one which corresponds to the absolute minimum of the $\chi^2$) is the correct solution of the problem. As discussed above, the comparison of radio and optical data 
reveals that the constant braking law assumption is a good, but not exact, approximation of the pulsar's phase development.
Taking this into account, the errors in Table~\ref{tab4} should be revised considering the range of values of the parameters obtained from all the fits that we consider acceptable. They are of the order of $10^{-4}$ for $\phi_0$, $10^{-9}$ s$^{-1}$ for $\nu_0$, $10^{-14}$ s$^{-2}$ for $\dot \nu_0$, $10^{-21}$ s$^{-3}$ for $\ddot \nu_0$, $10^{-1}$ for $n$, $10^{-12}$ s for $P$, $10^{-5}$ s for $t_{\rm arr}$. The errors inferred from the fits of the January and December runs separately (Table~\ref{tab3}) are more representative of the actual uncertainty for short (a few days) time intervals.


A consequence of the intrinsic irregularities of the pulsar rotation can be seen comparing the values of $\phi_0$, $\nu_0$ and $\dot \nu_0$ obtained from the `local' parabolic fits (Table~\ref{tab3}) and the global 2009 cubic fit (Table~\ref{tab4}). The differences in some of the fitted parameters (e.g. $\phi_0$ and $\nu_0$ in December) are larger than the uncertainty. This shows that a few observing nights are sufficient to reduce the statistical error to the point that we become sensitive to effects related to the long-term phase noise of the pulsar (see Section~\ref{sec6}). Despite these differences, the optical$-$radio phase shift observed in January and December is robustly recovered from both the `local' and global timing solutions, indicating that it is not an effect of radio data long-term averaging.

We studied also the distribution of the phase residuals about their best$-$fitting polynomials. The residuals are computed with respect to the timing solution reported in Table~\ref{tab4} (solution for $t_0=54846$). The optical phase residuals can be satisfactorily fitted with a Gaussian with a dispersion $\sigma$ between 15 and 22 $\mu$s (see Fig.~\ref{fig5}). The Gaussian distribution provides a good fit of residuals of single observing nights, as well as of residuals with respect to the 2009 coherent timing solution, with a $\sigma=17.3\mu$s, which is essentially the average value of single night residuals.

The width of the distribution is consistent with that induced by pure photon counting noise, as simulated using a synthetic signal that has the same pulse shape, average count rate, and background level of the Crab pulsar observed with Iqueye at NTT. 
We note that different observations have slightly different widths, most likely because of the different quality of the sky during the various nights, as discussed in Section~\ref{sec2}.

\section{Discussion and conclusions}
\label{sec6}

The measurement of the optical$-$radio phase shift, obtained with Iqueye in 2009 ($\sim 240 \, \mu$s in January; $\sim 160 \, \mu$s in December), is consistent with that determined in our previous measurement performed with Aqueye in Asiago in 2008 October ($\sim 230 \, \mu$s; \citealt{germana+12}). The optical pulse always leads the radio one. This is also in agreement with a number of other measurements performed after 1996 (see e.g. \citealt{oosterbroek+08}). Only a couple of previous measurements yielded different results (consistent with no or negative delay; \citealt{golden+00,romani+01}), possibly because some systematics were not adequately accounted for 
(e.g. the difference between the phase of the peak of the main pulse and the phase of the centroid of the main pulse, which is due to the asymmetric pulse shape). Therefore, although it is not possible to exclude a secular change of the phase delay, 
the available data show that the pulse in the optical leads that in the radio by 150-250$\mu$s (see also \citealt{strader+13}).
This can be caused by a different position of the two emitting regions (50$-$80 km), although this is difficult to understand as electrons close to the pulsar are highly relativistic and any travel time delay between electrons and photons is essentially negligible. Most probably, the optical and radio beams are misaligned (1.5$^0$$-$3$^0$) because at the position where electrons emit optical photons the magnetic field has a slightly different orientation with respect to that where radio emission takes place.


The counting statistics available in the optical band allows us to compute the timing solution with extreme accuracy from a small number of consecutive observing nights, with an error of only $\approx 3\times10^{-5}$ in fractional phase or $\approx 1 \, \mu$s in time (see Table~\ref{tab3}). 
The optical timing solution matches the radio phase quite closely, yet in addition to a presumably constant phase shift between optical and radio, we find small $\approx \ 10 \, \mu$s d$^{-1}$ phase drifts between the two solutions, which are consistent with timing noise deduced from radio and optical data and with residuals effects of changing dispersion measures on the radio data.

The distributions of phase residuals are well approximated by Gaussians. The best observations have a dispersion $\sigma \approx 15 \, \mu$s (with 2 s sampling), in agreement with theoretical expectations for pure photon noise-induced phase variability. However, a few January and December observations have a slightly larger spread ($\approx 20 \, \mu$s), which is most likely caused by worse quality of the sky.
We note that the dispersion of residuals in our data is much smaller than the dispersion of the residuals for Giant Radio Pulses (84 $\mu$s) recently reported by \citet{rameshbhat+13}.

Giving the exquisite precision of the Iqueye data, we were able to phase connect the January and December observing runs, obtaining a global third-order timing solution for year 2009 and an independent measurement of the pulsar braking index.
A comparison with the 2009 timing solution in the radio band shows that the pulsar makes small amplitude random excursions/oscillations around the ``average'' cubic spin-down (with residuals up to a few ms in phase and a jitter up to $\sim 10^{-8}$ Hz in frequency), that are statistically significant. Imposing that the number of optical turns be equal to that in the radio band, the optical and radio phases in January and December are very close, but at intermediate times phase predictions differ. As a consequence, the inferred values of the frequency, its derivatives and braking indices also differ. Therefore, while the third-order secular spin-down and braking index model are very good at reproducing the overall evolution of the phases (e.g. \citealt{groth75c}), an accurate description of the pulsar phase excursions/oscillations requires a frequent monitoring on a monthly time-scale. 

The optical observations reported here confirm the third-order secular spin-down and braking index model to a high degree of accuracy. During the period from 2009 January to December, the pulsar has followed a braking index law so closely that, from the two Iqueye sets of observations alone, we were able to count the almost nine hundred million turns that the pulsar has made with an uncertainty of only two turns.

Future and more regular observations taken with both Iqueye and its twin Aqueye, mounted at the Copernico telescope in Asiago, will offer an invaluable means to monitor the optical$-$radio phase delay, to understand the phase excursions around the braking index model and to pinpoint the deep physical origin of the pulsar spin-down mechanism.

\section*{Acknowledgements}

We thank the referee for their report with helpful comments that improved the paper, Claudio German\`a for useful discussions, and Ivan Capraro, Andrea Di Paola and Claudia Facchinetti for their contribution to the realization of Iqueye.
This work is based on observations made with ESO Telescopes at the La Silla Paranal Observatory under programme IDs 082.D-0382 and 084.D-0328(A). We acknowledge the use of the Crab pulsar radio ephemerides available
at the web site of the Jodrell Bank radio Observatory
(http://www.jb.man.ac.uk/$\sim$pulsar/crab.html; \citealt{lyne+93}).
This research has been partly supported by the University of Padova under the Quantum Future Strategic Project, by the Italian Ministry of University MIUR through the programme PRIN 2006 and by the Project of Excellence 2006
Fondazione CARIPARO.

\appendix

\section{Transformation of the $\psi$-phases between two different temporal gauges}
\label{app}

In the following, $t_{\rm P}(t)$ is the time of arrival of the main pulse at/after time $t$, while ($\hat P_{\rm init}$, $\hat t_0$) and ($P_{\rm init}$, $t_0$) refer to the two different temporal gauges. As mentioned above, we assume that the pulsar period changes slowly in time and, as a consequence, the two reference periods $\hat P_{\rm init}$ and $P_{\rm init}$ are not very different from the actual pulsar period and from each other ($|\hat P_{\rm init}$/$P_{\rm init} - 1| <$ phase bin). We define
\begin{eqnarray}
&& \hat \alpha (t) = \frac{t_{\rm P} - {\hat t_0}}{\hat P_{\rm init}} = \frac{t_{\rm P} - t_0}{\hat P_{\rm init}} + \frac{\Delta t}{\hat P_{\rm init}}
\label{alpha1} \\
&& \alpha (t) = \frac{t_{\rm P} - t_0}{P_{\rm init}}
\label{alpha2}
\end{eqnarray}
where $\Delta t = t_0 - {\hat t_0}$ and the fractional part of $\hat \alpha$ and $\alpha$ is measured as described in Section~\ref{sec3} (like the $\psi$ phase). As the time of arrival of a single pulse determined in the two different temporal gauges is equal, from the previous equations we obtain:
\begin{eqnarray}
&& \hat \alpha (t) = \frac{P_{\rm init}}{\hat P_{\rm init}} \, \alpha (t) + \frac{\Delta t}{\hat P_{\rm init}}
\label{gaugetrans1}
\end{eqnarray}
Considering that, modulo 1, it is $f\!r[\hat \psi]=f\!r[-\hat \alpha]$ and $f\!r[\psi]=f\!r[-\alpha]$, after some manipulations we get:
\begin{eqnarray}
f\!r[\hat \psi] + (n - {\hat n} - \hat \psi_n) = u \, f\!r[\psi] - \left( u - 1 \right) n - \frac{\Delta t}{\hat P_{\rm init}} - {\hat \psi_n}
\label{gaugetrans2}
\end{eqnarray}
where $n={\rm int}[(t_{\rm P} - t_0)/P_{\rm init}]$, $\hat n={\rm int}[(t_{\rm P} - {\hat t_0})/\hat P_{\rm init}]$, $u=P_{\rm init}/\hat P_{\rm init}$ and $\hat \psi_n$ is an integer shift needed to align phases between successive nights. Writing $\hat n={\rm int}[(t_{\rm P} - t_0)/\hat P_{\rm init}] + {\rm int}[\Delta t/\hat P_{\rm init}]$, we finally obtain:
\begin{eqnarray}
f\!r[\hat \psi] + (n - {\hat n}^{'} - \hat \psi_n) = u \, f\!r[\psi] - \left( u - 1 \right) n - v - {\hat \psi_n}
\label{gaugetrans3}
\end{eqnarray}
where $v=\Delta t/\hat P_{\rm init} - {\rm int}[\Delta t/\hat P_{\rm init}] = f\!r[\Delta t/\hat P_{\rm init}]$. 
The integer part of $\hat \psi$ measured from $t_0$ consists of two contributions, the integer difference between the two temporal gauges ($n - {\hat n}^{'}$) and the integer shift ($-\hat \psi_n$). Equation~(\ref{gaugetrans3}) corresponds to equations~(\ref{eqpsij}) and~(\ref{eqpsid}).

The same equation can be derived asking that the true rotational phase $\phi$ of the pulsar is gauge-independent. From equations~(\ref{phijandec1}) and~(\ref{phijandec2}), we have:
\begin{eqnarray}
&& \phi (t) = \hat \psi (t) + \frac{(t-\hat t_0)}{\hat P_{\rm init}} 
\label{phi1}  \\
&& \phi (t) = \psi (t) + \frac{(t-t_0)}{P_{\rm init}} + N \, ,
\label{phi2}
\end{eqnarray}
By equating these two expressions for $\phi$ and using the quantities defined above, after some manipulations we obtain:
\begin{eqnarray}
&& \hat \psi = f\!r[\psi] - (u - 1) \left( \frac{t-t_0}{P_{\rm init}} \right) - v + N^{'} \, ,
\label{phigauge1}
\end{eqnarray}
where $N^{'}$ is an integer. Adding and subtracting $u \, f\!r[\psi]$, we finally have:
\begin{eqnarray}
&& \hat \psi = u \, f\!r[\psi] - (u - 1) \left( \frac{t-t_0}{P_{\rm init}} + f\!r[\psi] \right) - v + N^{'} \, .
\label{phigauge2}
\end{eqnarray}
Since $f\!r[\psi]=f\!r[-\alpha]=-f\!r[(t_{\rm P}-t_0)/P_{\rm init}]$, equation~(\ref{phigauge2}) can be written as
\begin{eqnarray}
\hat \psi = u \, f\!r[\psi] - (u - 1) \left[ {\rm int}\left( \frac{t-t_0}{P_{\rm init}} \right) + f\!r\left[ \frac{t-t_{\rm P}}{P_{\rm init}} \right] \right] - v + N^{'} \, .
\label{phigauge3}
\end{eqnarray}
In the assumptions stated above, it is $|u-1| \ll 1$. So, we can neglect the second term in square brackets on the right hand side and equation~(\ref{phigauge3}) reduces to equation~(\ref{gaugetrans3}) (${\rm int}[(t-t_0)/P_{\rm init}]={\rm int}[(t_{\rm P}-t_0)/P_{\rm init}]=n$).

\bsp

\label{lastpage}

\end{document}